# HOW TO TRAIN YOUR SPACE TESTER, PART 2: ENABLING CAPABILITIES

Evelyn Kent[1], Jason Gabriel[2], Anna Gunn-Golkin[3], Charles Langdon[3], Michael Nayak[4], Emily Remeta[5], Michelle Willett[6]

1: Air Force Research Lab Space Vehicles Directorate
2: Department of the Air Force Rapid Capabilities Office
3: 3d Test and Evaluation Squadron
4: DARPA Defense Sciences Office
5: 20th Space Control Squadron
6: Information Dominance Directorate, SAF/AQ

## ABSTRACT

The need for space test professionals is growing rapidly, due to both the establishment of the US Space Force (USSF) and the extremely rapid growth of the commercial space industry. As there is increasing need to deliver capabilities to both warfighters and commercial customers - in, from, and through space - the future of space test looks bright and complex. A better training foundation is needed to impart test-mindedness to all space professionals to enable this future. Discussed from a military perspective, but applicable across the space industry, there are three critical aspects of foundational training:

(1) Realistic trainers and simulators: Representative simulators are the safest way to gain repetitions in dynamic decision-making, satellite envelope expansion, fault response, and to develop ways to operate through a contested, degraded, and operationally limited (CDO) environment. For warfighters, this extends to tactics employment in an increasingly congested domain. This paper introduces a new term encompassing the variables of space maneuver (GET CLASS) as a standard for Guardian (USSF member) training simulators.
(2) Interoperable Software Tools: As technology progresses and programs make independent decisions, Guardians must continuously adapt to new software tools without reliance on a checklist. Common standards and interfaces must be developed and enforced in the interest of informing a common operating picture of the remote space environment, and, crucially, to equip a space test range to host a variety of platforms.
(3) Cross-functional networking: Operations, test, engineering, and program management communities must work hand-in-hand to deliver effective capabilities. Blending career fields and mission areas inspires innovation and propagates improved capabilities across the space industry.

# INTRODUCTION

Two years ago, the authors discussed four big-picture challenges facing space test training: testing beyond imposed, strict operating limits; expanding permitted procedures at the operator's disposal; space-domain specific safety nets, and qualification and currency considerations[1]. These articles were published at a relevant time to inform the US Air Force Test Pilot School's (USAF TPS) Space Test Fundamentals (STF) course, which executed for the first time in January 2021. Following the success of STF, USSF and Air Force Test Center leadership jointly agreed to expand STF into a parallel course alongside the aircraft flight test track at USAF TPS, partnering on academics and test events with the flight test students, and changing its name to USAF TPS Space Test Course (STC). STC currently provides hands-on training in test fundamentals, systems test, space science application, and advanced space system test and evaluation.[2] In March 2022, the USSF published its Space Test Enterprise Vision[3] which calls on the service to deliver "credible warfighting capabilities on operationally relevant timelines" and "broad foundational test training for all of the test workforce."[4] We turn our attention to this broad force-wide training in this follow-up work, with an eye to the future of space test.

As the STC curriculum is refined and the cadre of USSF space testers grows, the next glaring need is for the larger percentage of USSF Guardians to be trained in test-mindedness from the beginning of their careers. Test-mindedness entails critical thinking, deep systems understanding, and a focus on learning, which is a pivot from the historical checklist-driven culture that relied heavily upon development contractors for troubleshooting and technical expertise. Previous cultural norms across the US Air Force, from which many early space professionals emerged, further bred this mentality of devoutly following checklist items and rewarding procedural excellence. In his review of *Organizational Disfunction in the US Air Force*, Lt Col Bud Fujii-Takamoto describes the potential consequences levied on Nuclear Missile Operations Officers who failed to pass written tests or otherwise demonstrate checklist perfection: "failure to meet any of these milestones could mean failure to promote and the end of a career … paradoxically, skill in the primary occupational specialty resulted in the 'reward' of administrative duties, facilitating the removal of expertise from the alert crews to the back offices."[5] Lt Col William Sanders further illustrates Air Force Space Command's journey to procedural excellence, which was needed in the 1990s to consistently operate technically-challenging exquisite systems. "Space systems outperformed expectations and contributed to tactical operations on the largest scale in human history. Yet neither the processes nor the personnel were optimally integrated into combat operations. … The warfighters recognize

---

[1] Nayak, et al. "How to Train your Space Tester: Big Picture Challenges facing Space Test Training." 2020, 2021.
[2] USAF press release (Aug 2021), "USAF TPS graduates first 'production' Space Test Fundamentals class."
https://www.edwards.af.mil/News/Article/2742377/usaftps-graduates-first-production-space-test-fundamentals-class/
[3] Available online at https://www.spaceforce.mil/Portals/1/Documents/Space-Test-Vision.pdf
[4] Thompson, "Space Test Enterprise Vision," March 2022.
[5] Fujii-Takamoto, "Organizational Dysfunction in The Us Air Force: Lessons from The ICBM Community": 44, June 2016

mere compliance and rigid procedures are not sufficient to compete with a thinking adversary in a dynamic environment."[6]

Most authors of this work attended the inaugural offering of the space test track of USAF TPS; others are graduates of the air test track with experience in both high-risk experimental aircraft and prototype space vehicles. STC attendees hailed from the operations, test, engineering, and program management communities. As part of the course, this diverse group experienced and discussed the tools and trainers currently employed by the USSF, providing the basis for the perspectives contained herein. As in the USAF flight test community, STC graduates will be only a fraction of the larger USSF test workforce. However, the principles outlined in this paper are applicable to the quickly-growing space test community writ large. We propose three focus areas for foundational Guardian training to emphasize test-mindedness: trainers and simulators, interoperable software tools, and cross-organizational networking.

## TRAINERS AND SIMULATORS
*"All models are wrong, some are useful."*
*George E. P. Box*

For decades, the flight test community in the air domain has understood the need for, and invested in the development and availability of, high-fidelity simulators. These simulators provide a safe environment for training pilots and test teams, practicing and validating flight procedures, developing flight test techniques, and predicting aircraft responses. Test pilots and engineers may spend considerable time in high fidelity simulators prior to a sortie. Test points "for score" may even be accomplished via a sufficiently validated simulator. From a safety standpoint, simulator rehearsals enable the test team to identify, anticipate, and prepare for knock-it-off criteria and establish maneuver setup conditions to minimize risk. In cases where simulator fidelity is not fully sufficient for test rehearsals, aircraft test teams almost always have an option to conservatively approach a test point, land if there is an unexpected test event, service the aircraft, refuel, and continue the test campaign after analysis or implementing a fix.

Space test requires high-fidelity simulators for the same purposes. Trainers allow Guardians to interact with representative control software to become familiar with both normal and contingency operations. In his Operational Test and Training Infrastructure guidance, General David Thompson identified the capability gaps between today's trainers that are focused on "procedural currency," and more robust trainers and simulators that are needed for realistic results.[7] While useful, today's spacecraft simulators focus on nominal operations in a benign environment. As such, they do not replicate external factors in the highly-dynamic space environment in which Guardians must now perform. Guardians must protect and defend spacecraft in a CDO environment, and in the future may need to test and execute on-orbit servicing and even manufacturing capabilities. Furthermore, space testers rely more heavily on simulators for test safety in a maintenance-limited domain. For example, the lack of refueling opportunities in space imposes a high cost on test maneuvers; therefore comprehensive test campaigns may use a minimal set of test points, then use the results from those test points to validate a simulation in which to execute the rest of that campaign. Guardians require simulators that go beyond procedures, allowing them to truly train and test as they will fly.

---

[6] Sanders, "Space Force Culture", June 2022.
[7] Thompson, David D. "USSF OTTI," June 2022.

An extra layer of realism is required to train and prepare Guardians for dynamic space operations throughout all levels of Guardian training. We shall use rendezvous and proximity operations (RPO) as an example of a highly dynamic multi-spacecraft test environment to make our case, analogous to formation flying in airborne flight test. An RPO maneuver can involve a "target" spacecraft, which represents the rendezvous point, and a "chase" spacecraft, which is the vehicle primarily maneuvering. The space shuttle orbiter was an example of an RPO chase spacecraft. The orbiter would maneuver towards and then dock with its cooperative target spacecraft, the International Space Station (ISS). While crewed space operations are an easily recognizable example, RPO applications are growing in both civil[8] and commercial[9] space, as on-orbit servicing and manufacturing opportunities increase. In military space, the USSF must field robust simulation capabilities to prepare Guardians for this dynamic mission set, among others. USSF trainers, for both fundamental Guardian training and capability validation, must dynamically incorporate all aspects of an orbital engagement: geometry and time of flight (TOF), as well as concerns related to external coordination, communication links, and on-board systems and subsystems.

We refer to these considerations by a new term, "GET CLASS" – GEometry, TOF, Coordination, Links, and SystemS. Simulators that include all GET CLASS variables provide the best training for Guardians who need to operate in the highly dynamic environment described above. Guardians should understand each of the GET CLASS terms and their significance, as described below.

**GET CLASS – GEometry**
*Geometry* involves all characteristics of the orbital regime and surrounding elements: time-space-position information on the satellite, forecast of future position, and external environmental effects such as atmospheric drag and solar radiation. A space trainer for RPO missions must also be able to represent two or more spacecraft's relative geometry, defined by relative position and velocity between spacecraft. Guardians must frequently develop maneuvers for RPO or in response to collision risks with surrounding satellites or debris, which requires initial position and velocity conditions and an appropriate orbital dynamics propagation engine.

**GET CLASS – TOF**
Similarly, a trainer should allow the user to vary the *TOF*, which is the time it takes a spacecraft to move from one point in orbit to another. During RPO, fuel is expended in a velocity change ($\Delta V$) to affect the distance from another spacecraft. During maneuvers, $\Delta V$ and TOF are typically inversely proportional; the more energy is added to an object's orbit, the more quickly it reaches its destination. Space test measures of performance, for example, might include minimized TOF to validate an operational requirement of arriving at its destination rapidly, or minimized $\Delta V$ to conserve limited fuel. Many trainers already model and provide adequate training opportunities for the "GE-T" components of GET CLASS, but must not compromise these components to add the "CLASS" components described next.

---

[8] For example: NASA/Maxar OSAM-1 (formerly Restore-L)
[9] For example: Lockheed-Martin/Orbit Fab RAFTI, Northrop/SpaceLogistics MEV series (fueled Intelsat)

**GET CLASS – Coordination**
Dependencies on external requirements for space operations can be captured through a robust simulation of *coordination* processes. On-orbit space test is not done in a vacuum. Support from antenna networks to provide the links described below, cooperation with operators of other spacecraft, and observation data from space domain awareness (SDA) providers are all critical to executing a safe, secure, and responsible on-orbit test. These areas of support must be requested through specific processes, and fundamental Guardian training should include making these requests appropriately. SDA observations are critical to an RPO test campaign in particular; before a spacecraft can close the loop to navigate in proximity to another spacecraft, it must be cued from an external source to facilitate the initial rendezvous. For commercial collision avoidance in the future, perhaps facilitated through a future space traffic control network[10], closing maneuvers observed must be shared quickly with the responding spacecraft operator(s) so they can appropriately update their maneuver plan. Internal to the operations floor, operators should use a standard voice protocol to clearly and concisely communicate while conducting a high-stakes objective. Finally, given the nature of interconnected space, ground, and link segments that are frequently controlled by different organizations, clearly defining the test boundaries (i.e. which portions or segments are "under test" and which are supporting or witnessing sensors, and therefore which organizations will be involved) is a key element of the pre-test coordination process.

**GET CLASS – Links**
Communication *links* from the ground segment to most spacecraft are not constant or guaranteed, and they are often cumbersome to schedule and maintain throughout the course of a test. While voice, video, and telemetry links are "just another" test resource in the flight test community like range time or an instrumentation pod, space test requires active, continuous management of command and control (C2) links to enable a successful test. For military operators, contact time must be scheduled in advance through the Satellite Control Network (SCN) or another antenna network. SCN time is a particularly limited resource servicing a multitude of operational military spacecraft, and test needs will frequently be out-prioritized by national security needs. Even for commercial operators fielding their own networks, contact times can be a limited resource, particularly with growing constellation sizes or limiting orbital geometry. Links also have latency and limited data rates that can impact test accomplishment and results. A trainer must therefore replicate realistic link scheduling and management. Space testers executing RPO, for example, must be cognizant of which dynamic or elevated risk maneuvers require continuous C2 links, as well as times when losing a link would require a "knock-it-off" call to stop the test. Additional discussion about safety nets to mitigate impacts may be found in the precursor work.[11]

---

[10] Johnson, Nicholas L. "Space traffic management concepts and practices." Acta Astronautica 55.3-9 (2004): 803-809.

**GET CLASS – SystemS**
Finally, a robust trainer must also replicate on-board spacecraft *systems*. For some space missions, systems may include the primary payload, such as the camera on a commercial imagery spacecraft, as well as supporting subsystems such as electrical power generation and thrusters for orbital maneuvering. For our RPO example, onboard sensors facilitate navigation in proximity to another spacecraft, but the success of an RPO mission also relies upon the thrusters, attitude control system, and solar power generation, all of which must be accurately modeled in the simulator and trainer to provide effective training or representative test execution.

Robustly simulating the entirety of GET CLASS allows space testers to embrace a large scope of their responsibility with respect to dynamic test campaigns. A truly effective space trainer will include additional variables specific to a given mission or satellite, overlaid with GET CLASS information, to provide the best test and training value.

In the bustling and contested space environment, operators of all spacecraft should expect their concept of operations (CONOPS) to expand over mission life, requiring procedures not developed during the developmental test campaign. This increased CONOPS is analogous to expanding a performance envelope in the flight test world. Any new procedures or techniques should be validated and practiced on a mission-specific and flight-representative simulator, prior to executing on-orbit. Post-launch, real vehicle data feedback should inform trainers and simulators to increase the accuracy of simulated procedure and scenario results; the live range will regularly provide such data for enhancement of simulators and models.[12] Using GET CLASS trainers to practice tasks and procedures, and the importance of validation with flight data, should be standard practice trained into every Guardian from the beginning of their career. Furthermore, each USSF operations center should include a flight-representative mission simulator for both proficiency training and envelope expansion.

As a final touch, simulators may include access to operational software tools. Including software tools in training scenarios will help familiarize Guardians with these resources from the start of their careers. Important considerations for software tools are discussed in the next section.

## INTEROPERABLE SOFTWARE TOOLS
*"It is a capital mistake to theorize before one has data."*
*Sherlock Holmes*

---

[11] Nayak et al., 2020, 2021
[12] Thompson, David D. "OTTI," June 2022.

Space professionals need to see beyond their system level to the whole picture, including ground-based, space-based, and environmental data points. Rather than today's paradigm of completing tasks by checklist, Guardians of the future must be masters of their tools and able adapt to new ones without total reliance on checklists for critical decision-making. There are a growing number of commercial, government, and open-source tools that can help an operator accomplish common space mission functions, including C2, resource scheduling, SDA, and mission planning. With the vast menu of software options, Guardians must be adaptable to new software tools, which should be designed to be intuitive and to not require a procedure for common functions as they do today. To keep the market truly open and future-proof USSF systems, Guardians acquiring tools must collectively enforce common standards and interfaces so that mission data may be integrated into a common operating picture (COP). Common standards and interfaces are especially crucial for the test and training range that will support diverse missions and assets, enabling the integration, adaptability, collaboration, and interoperability that General Thompson envisions.[13] Foundational Guardian training should include orientation and use of a variety of tools and methods to build critical thinking skills and make Guardians more adaptable to changing situations. Two key tool categories are space domain awareness and mission-enablers.

In the first category, SDA databases of all registered spacecraft, often updated by ground-based radar, exist to provide a baseline picture of everything on orbit. Spacecraft state (position and velocity) information from these databases can be integrated into simulators or other tools to plan and predict outcomes for objectives like RPO, orbital adjustments, launch insertion, and collision risk assessments. Space weather prediction tools can be overlaid on state information to inform space operators of periods of increased likelihood of charging events. Space-to-ground communication link budget calculation tools estimate signal strength based on geometry and radio parameters to identify best contact opportunities. For space test, these tools can add realism to developmental test, and provide real world threats and challenges for operational test. Guardians should be trained on and regularly exposed to these tools to accomplish mission objectives effectively.

The other category of common tools are mission-enabling tools that include C2, scheduling, and planning. For test purposes, a simple open-source tool may prove effective for engineers that need quick access to raw data. For long-duration operational missions, customized mission functions and an enhanced user experience are worth a larger investment into software development. A space test range will need to procure its own assets that might run on inexpensive and common software, but must also be able to receive and interact with residual capabilities that have mission unique software. A hybrid environment like this will rely on strict interfaces and standards to ensure successful interoperability, and the operators in such a hybrid environment must be comfortable with constantly learning unfamiliar tools.

---

[13] Thompson, David D. "OTTI," June 2022.

Approved for public release; distribution is unlimited. Public Affairs release approval #AFRL-2022-3716

As we move into the future of space test, Guardians' tools must be increasingly interconnected. An interface control document (ICD) provides details of the inputs and outputs of a tool to provide an easy interface with other tools or users – ideally defined by publicly available open architecture standards. It is our belief that the application of data standards and common interfaces will facilitate data sharing across the space community, which can be used to build a COP and to better coordinate multi-mission operations.

In a future of global interconnectedness, there are two limitations of software tools to note. First, while open-source software provides a cheap and easy avenue to gather data, Guardians must be aware of the software's vulnerabilities. USSF program managers must budget for security evaluation of their software.[14] Second, Guardians must be wary of the possibility of software latency or outages. Anecdotally, data manipulation tools always seem to lag or fail during time critical operations, leaving operators without the tools they've come to rely on during training. But users need to be prepared to experience latency and mission plan accordingly. This issue can be replicated in training by limiting tool usage, or removing access, so that Guardians receive practice in moving to manual backup capabilities.

Foundational Guardian training must shift focus from today's paradigm of rote practice in a single mission-unique tool, to exposure to several common tools and discussion of the strengths and limitations of each. The curriculum should prioritize development of critical-thinking skills rather than adherence to a checklist. Unlike procedural knowledge, intimate familiarity with tools results in better understanding of the domain, adaptability to the inevitably continuous flow of new tools, and increased confidence in time-critical situations. USSF has publicized their vision for skills-based hiring and capabilities-based crew positions in its Guardian Ideal, and test professionals will surely be a key component of their talent management strategy.[15]

Finally, cross-talk and career flow between program managers, engineers, operators, and testers, and – at a larger scale – between military, civil and commercial space test, will ensure that Guardians have knowledge of and access to the advances in these tools.

## NETWORKING
*"If I have seen further, it is by standing on the shoulders of giants."*
*Sir Isaac Newton*

Success in a challenging, high-stakes field like test will always require communication across the community to share lessons learned, improve upon past mistakes, and prevent duplication of effort. As the space industry grows larger and more diverse, with competing goals ranging from protection to commercialization to human spaceflight, open channels of communication and forums to discuss training methods and data standards will grow in importance, at levels from broad to narrow audiences. Additionally, the results of a single test campaign could inform or significantly decrease costs for others' test plans, and the test community will also need to reach outward to support the joint force. As a burgeoning community, space testers have a unique opportunity to both capitalize on inspiration from existing networking constructs and establish their own.

---

[14] Tucker, Patrick. "Space Runs on Open Source Software. The US Air Force is Fine with That." 11 July 2022. https://www.defenseone.com/technology/2022/07/space-runs-open-source-software-us-air-force-fine/374103/
[15] US Space Force. "The Guardian Ideal":13, September 2021. https://www.airforcemag.com/app/uploads/2021/09/21SEPT-USSF-GUARDIAN-IDEAL.pdf

Professional societies such as the Society of Flight Test Engineers (SFTE) provide diverse perspective, from the international to the regional level, and across government to commercial flight test. This interaction permits the opportunity to network with the larger community in a domain-agnostic manner to improve and maintain test fundamentals, stay abreast of the latest methods and standards, and consider ways to adapt innovative methods being adopted by one domain into another. Space testers should engage at this level to share unique space test challenges and accomplishments.

From there, more tailored audiences may be desired to discuss domain-specific test. The military air test community holds an annual Developmental Test Working Group (DTWG) to discuss test successes and failures, lessons learned, and upcoming events. Perhaps most importantly, this forum presents an opportunity to make peer-to-peer connections that can be invaluable for informal feedback on test methodologies, common errors, and ways to improve fidelity. Recently, in early 2022, USSF's office of test and evaluation held its first Integrated Test Working Group (ITWG). The goals of this event were analogous to that of DTWG; this continuing annual meeting of testers, at appropriate classification levels, will be crucial to maintaining the network of space testers, forging new partnerships, and developing or discovering domain-specific methods or tools for test.

Finally, networking on a person-to-person level can be just as important to help young testers find mentorship and become confident decision-makers. TPS has an excellent start on building a peer network. Students are selected from a variety of backgrounds and career fields, which provides a much-needed diversity of perspectives. However, this is just the start: every space tester should seek out continuous learning from their peers. Simply "knowing who to call" when uncertain of a way forward can be a huge benefit, and enhance collaboration across networks, organizations, and platforms. Individual networking can also provide young test professionals insight about future opportunities.

At any of these levels of interaction, an unknown peer may emerge from the audience to provide insight into a useful capability or resource. A planned test may have been accomplished in part by another program, or an architecture problem may have already been solved. In the government, leveraging an existing contract can be easier and faster than seeking to draw up a new contract for a similar capability. Socializing ideas among peers is an age-old method of refining those ideas by fire.

On the training side, networking external to the space community is just as important as interactions within the space community. A great example of this external networking is the recurring two-week Space Flag exercise that most recently concluded on August 19, 2022. "As a new branch of the military, the Space Force…has to learn how to integrate with the other armed services during an actual operation, and that's a key part of the training."[16] This exercise, and others like it, provide the opportunity for Guardians to get to know the broader community, understand the importance of their mission, and practice using the communication channels that will be required outside of training.

For Guardians, collaboration between acquisitions and operations is becoming a programmatic requirement once an acquisition reaches its developmental test campaign. USSF has prescribed Integrated Test Forces (ITFs), bringing acquisitions, test, and operations personnel together to conduct a comprehensive integrated test campaign. In this tri-lateral group, testers will bring test rigor, best practices, resources, and expertise to maximize test effectiveness. Acquirers bring enabling contracts and program management expertise, and especially at the start of a test campaign have a greater familiarity with the space vehicle and its design than anyone else on the team. Operators provide invaluable user input to create a truly sustainable capability.

While the Combined Test Force (CTF) concept is not new to air test, the ITF paradigm will challenge existing cultural barriers in military space test. "The operators' rigor and a penchant for standards are not, in themselves, bad things, but the Space Force must balance checklist discipline with creativity and innovation."[17] Test units must come in with a collaborative attitude; an independent test organization that values the interests of the end user. Operators must exercise patience with the acquisition community's regulations and processes; compromises may need to be struck to meet cost or schedule constraints. Finally, acquisition units that historically either take ownership or relinquish control to a test unit must strike a balance of participation in the test campaign, allowing their test and operations partners to advise them on intelligent investment in resources to produce a long-lasting capability. In other words, fulfilling the vision of "intentional workforce crossflow between acquisition, test, and operations"[18] in a service-wide approach to test.

## CONCLUSION

In this follow-on to our seminal work on training space testers, we have shifted focus from mid-career training of space testers to early-career training for all Guardians. The future of the space domain, while bright, requires a change to today's foundational Guardian training to instill better test-mindedness overall. As STC continues to feed the pool of space test cadre, the USSF must turn its attention to developing all Guardians with critical thinking and adaptability. In our experience, the three critical components of foundational Guardian training to meet that goal are:

(1) Realistic trainers and simulators: More realistic, flight-representative simulators capable of training Guardians for highly dynamic on-orbit campaigns are required. GEometry, TOF, Coordination, Links, and SytemS (GET CLASS) considerations improve the realism of a trainer or simulator.
(2) Interoperable Software Tools: The distant nature of spacecraft can complicate situational awareness for both commercial entities and warfighters. Integrating data from interoperable software tools across the force, and making that data available to all users of the domain, provides a coherent common operating picture.
(3) Cross-functional networking: Diverse communities must mingle to deliver effective capabilities, and to create a tight-knit network that transcends the bounds of organization and career specialty.

---

[16] Erwin, Sandra. "Space Force wargame challenges satellite operators to think critically." 21 Aug 2022. https://spacenews.com/space-force-wargame-challenges-satellite-operators-to-think-critically/
[17] Sanders 2022.
[18] Thompson, David D. "Space Test Enterprise Vision," March 2022

## ACRONYMS, ABBREVIATIONS, SYMBOLS

C2: Command and Control
CDO: Contested, Degraded, and Operationally Limited
CONOPS: Concept of Operations
COP: Common Operating Picture
CTF: Combined Test Force
Delta V, $\Delta V$: Change in Velocity
DTWG: Developmental Test Working Group, gathering of air test TPS graduates
GET CLASS: New term encompassing important aspects of a high fidelity space system simulator, GEometry, TOF, Coordination, Links, and SystemS
ICD: Interface Control Document
ISS: International Space Station
ITF: Integrated Test Force
ITWG: Integrated Test Working Group, gathering of space test TPS graduates
RPO: Rendezvous and Proximity Operations
SFTE: Society of Flight Test Engineers
SCN: Satellite Control Network
SDA: Space Domain Awareness
STC: Space Test Course
TOF: Time Of Flight
USAF TPS: United States Air Force Test Pilot School
USSF: United States Space Force

## ACKNOWLEDGEMENTS

The authors would like to thank the instructors and leaders who have been instrumental in developing STC and bit by bit ensuring test rigor in the developing Space Test Enterprise, especially Dr. Andrew "Sonic" Freeborn, Col Sebrina Pabon, and Col T. Nick Hague.

## BIOGRAPHIES

**Evelyn Kent** leads the Operations Cadre for the Space Vehicles Directorate, Air Force Research Laboratory. In addition to serving as the Lead Flight Director and Operations Lead for the largest unmanned structure in orbit, DSX, she has logged space vehicle operations time with two FalconSats, a number of National Reconnaissance Office (NRO) vehicles, and five AFRL experimental satellites (and counting). She is a graduate of the inaugural class of the USAF Test Pilot School's Space Test Fundamentals Course.

**Jason Gabriel** is a Program Manager and Astronautical Engineer in the United States Space Force, leading system acquisitions and integrated test of multiple space programs. His background includes Program Management for the Department of the Air Force, satellite operations for the National Reconnaissance Office, scientific research on inertial navigation, and foundational studies on orbital debris mitigation. He holds a civil commercial pilot certificate with over 500 hours in airplanes, helicopters, gliders, and balloons (although he hasn't flown a balloon to space … yet). He is a graduate of the inaugural class of the USAF Test Pilot School's Space Test Course.

**Anna Gunn-Golkin** is the Commander, 3d Test and Evaluation Squadron, US Space Force. She was a guest instructor at the inaugural USAF Test Pilot School's Space Test Fundamentals Course.

**Charles Langdon** is currently a test director in the 3d Test and Evaluation squadron, US Space Force. His experience began in satellite system operations for two years for military satellite communications. Then proceeded to mission plan for another three years for experimental space systems. He is a graduate of the inaugural class of the USAF Test Pilot School's Space Test Fundamentals Course.

**Michael Nayak** (member, SFTE) is a Program Manager with the Defense Sciences Office, DARPA. His experience spans work as a space shuttle engineer; flight director for multiple experimental spacecraft; a skydiving instructor; a planetary scientist at NASA Ames; research section chief for the DoD's largest telescope; instructor flight test engineer and instructor pilot. He is a USAF Test Pilot School graduate, Rotary National Award for Space Achievement recipient, and has 1,000+ hours of flight time in 40+ aircraft including the F-16, T-38, EA500 and BE-76.

**Emily Remeta** is the Weapons and Tactics Flight Commander at the 20th Space Surveillance Squadron for Space Delta 2. Her experience spans three ground-based missile warning, missile defense and space domain awareness radars. She has worked with MIT Lincoln Labs to develop operational tactics and is an initial member of the USAF MIT Artificial Intelligence Cohort. She is a graduate of the inaugural class of the USAF Test Pilot School's Space Test Fundamental Course.

**Michelle Willett** is a Program Element Monitor and perhaps one of the Air Force's last remaining space geeks. Her experience spans three warfighting domains and includes flight test, software systems engineering, and space policy. She is a graduate of the inaugural class of the USAF Test Pilot School's Space Test Fundamentals Course.